\documentclass[article,preprint]{revtex4}
\usepackage{amssymb}

\usepackage{graphicx}
\usepackage{amsmath}

\begin{document}

\title{Irradiated asymmetric Friedmann branes}
\author{L\'{a}szl\'{o} \'{A}. Gergely and Zolt\'{a}n Keresztes}
\affiliation{Departments of Theoretical and Experimental Physics, University of Szeged,
6720 Szeged, D\'{o}m t\'{e}r 9, Hungary}

\begin{abstract}
We consider a Friedmann brane moving in a bulk impregnated by radiation. The
setup is strongly asymmetric, with only one black hole in the bulk. The
radiation emitted by this bulk black hole can be reflected, absorbed or
transmitted through the brane. Radiation pressure accelerates the brane,
behaving as dark energy. Absorption however generates a competing effect:
the brane becomes heavier and gravitational attraction increases. We analyse
the model numerically, assuming a total absorbtion on the brane for $k=1$.
We conclude that due to the two competing effects, in this asymmetric
scenario the Hawking radiation from the bulk black hole is not able to
change the recollapsing fate of this brane-world universe. We show that for 
\textit{light} branes and early times the radiation pressure is the dominant
effect. In contrast, for \textit{heavy} branes the self-gravity of the
absorbed radiation is a much stronger effect. We find the \textit{critical}
value of the initial energy density for which these two effects roughly
cancel each other.
\end{abstract}

\email{gergely@physx.u-szeged.hu\\
zkeresztes@titan.physx.u-szeged.hu}
\date{\today }
\maketitle

\section{Introduction}

According to brane-world models, our universe is a hypersurface (the brane)
embedded in a $5$-dimensional spacetime (the bulk), in which Einstein's
gravity holds. The attempts to increase the number of dimensions in physics
have a long history, originating with the pioneering works of Kaluza and
Klein. In the more recent models presented in \cite{ADD}, the (flat) extra
dimensions are still compactified. The novelty of the brane-world models
generalizing the original proposal \cite{RS2} of Randall and Sundrum (RS)
consist in allowing for one additional \textit{noncompact} dimension.
Instead of compactifying, the bulk is warped.

Gravitational dynamics on the brane is given by a modified Einstein
equation, which arises from the projection of the 5-dimensional dynamics and
the junction conditions across the brane. A major difficulty in this theory
is that due to a new source term, arising from the projection of the bulk
Weyl tensor to the brane, the system of differential equations describing
gravitational evolution is not closed on the brane. This new gravitational
theory allows general relativity as the low energy density limit (as
compared to the brane tension $\lambda $).

Brane-world models were initially $Z_{2}$-symmetric, with identical copies
of the bulk on the two sides of the brane. Equivalently, the brane could be
imagined as a domain boundary. This assumption, although it considerably
simplifies the model, seems unnecessarily restrictive in the context of
brane-world models with curved branes (generalized RS type 2 models). In
this context, our observable universe may be imagined as a Friedmann brane
moving either in a Schwarzschild-anti de Sitter bulk (if the bulk is
vacuum), or in a 5-dimensional Vaidya-anti de Sitter (VAdS5) spacetime (if
the bulk contains radiation). The modified Einstein equation in the $Z_{2}$%
-symmetric case was given covariantly in \cite{SMS}, and generalized for
asymmetric embedding in \cite{Decomp}. The effects of asymmetric embedding
were also analyzed in a more generic class of models, containing induced
gravity contributions \cite{Induced}. There it was shown that the late time
universe behaves differently under the introduction of a small asymmetry of
the embedding: on the two branches of such theories. While on the RS branch
asymmetry produces late-time acceleration, on the DGP branch the late-time
acceleration is diminished by asymmetry.

Whenever there is radiation in the bulk, it can arise from both brane and
bulk sources. The situation when the brane radiates into the bulk was
already considered both in a symmetric \cite{LSR} and in an asymmetric
scenario \cite{RadBrane}, \cite{VernonJennings} in the framework of
generalized RS theories. In both scenarios a radial emission of radiation
into the bulk was considered. A more realistic set-up, allowing for the
emitted gravitons to follow geodesic paths, was advanced recently in \cite%
{Langlois}.

In this paper we discuss the other possibility, when the radiation is
emitted by sources in the bulk. For this, we embed the Friedmann brane
asymmetrically into VAdS5 spacetime. The symmetry of the embedding is
severely broken by allowing for a radiating black hole\ on one side and no
black hole on the other side of the brane. Due to this asymmetry, there is a
radiation pressure acting on one side of the brane. This radiation pressure
continuously accelerates the motion of the brane, leading to accelerated
cosmic expansion. In principle, this mechanism can be a \textit{dark energy
candidate}.

In brane-world models, standard matter is confined to the brane. Therefore
the radiation coming from the bulk should be of non-standard nature. We
discuss here the case when this is the Hawking radiation of the single bulk
black hole. The expression of the Hawking radiation in a 5-dimensional
Schwarzschild-anti de Sitter (SAdS5) spacetime was derived for the curvature
index $k=1$ in \cite{EHM}-\cite{GCL}. This energy density was employed, but
for $k=0$ in \cite{Jennings} for the study of asymmetrically embedded
Friedmann branes into VAdS5.

\begin{figure}[tbp]
\begin{center}
\includegraphics[height=7cm]{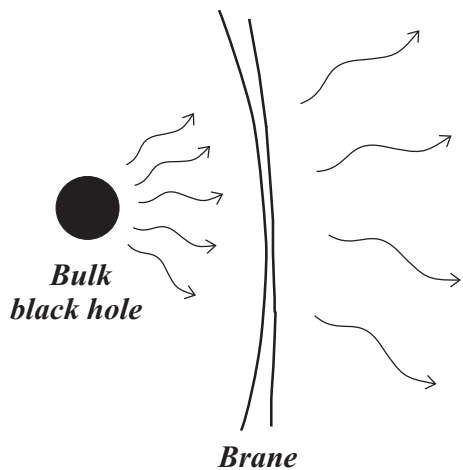}
\end{center}
\caption{The radiation from the left bulk black hole is partially absorbed
on the brane and partially transmitted through into the other bulk region
containing no black hole.}
\label{Fig:1}
\end{figure}

Some part of the radiation reaching the brane will be absorbed, some part
will be reflected and the remaining part will go through. We will study here
only the models without reflection in order to have the VAdS5 spacetime in
both bulk regions (Fig \ref{Fig:1}). We disregard the possibility of
reflection, because there is no known exact solution with cosmological
constant describing a crossflow of radiation streams. Such a spacetime with
both incoming and outgoing null dust streams, but without a cosmological
constant is known in $4$-dimensions \cite{GergelyND}, however no
generalization including a cosmological constant has been found. Thus we
discuss the case when the reflected component can be neglected.

We keep however the absorption, as an essential element of our model, in
contrast with the one presented in \cite{Jennings}, where there is a black
hole on each side of the brane, but their radiation is \textit{completely
transmitted} through the brane.

By retaining the absorption on the brane, we obtain novel features. The
absorbed radiation appears on the brane as continuously emerging energy.
Thus the energy density on the brane increases, strengthening the
gravitational self-attraction of our universe. As consequence of absorption,
cosmic expansion slows down.

Therefore there are two competing effects in our model: radiation pressure
accelerates, while the absorbed radiation decelerates cosmic expansion. We
would like to study how their balance affects the Friedmann brane-world. We
derive the relevant equations of the model in Section II and we comment on
the physical interpretation in Section III.

As our model emphasizes the effects of the absorption (as opposed to \cite%
{Jennings}), we finally choose a \textit{total absorption} on the brane in
Section IV. We also employ $k=1$, as the formula for the energy density of
the Hawking radiation was derived for this case \cite{EHM}-\cite{GCL}. Then
we introduce new dimensionless variables, adapted to our choice of $k=1$.

We provide quantitative results by numerical analysis in Section V. First we
discuss cosmological evolution in the case of a non-radiating bulk black
hole. Then, by switching on the radiation, we see how the two competing
effects modulate the cosmological evolution. We show that a critical
behaviour can appear, when these two competing effects roughly cancel each
other. We discuss the relevance of these results in a broader context in
Section VI.

Throughout the paper a tilde distinguishes the quantities defined on the
5-dimensional spacetime and a hat denotes dimensionless quantities and units
with $c=1=\hbar $ are chosen.

\section{Friedmann brane in Vaidya-Anti de Sitter bulk}

The Friedmann brane representing our observable universe is embedded in the
VAdS5 spacetime: 
\begin{align}
d\widetilde{s}^{2}& =-f\left( v,r;k\right) dv^{2}-2dvdr  \notag \\
& +r^{2}\left[ d\chi ^{2}+\mathcal{H}^{2}\left( \chi ;k\right) \left(
d\theta ^{2}+\sin ^{2}\theta d\phi ^{2}\right) \right] \ ,  \label{ChVAdS5}
\end{align}%
where 
\begin{equation}
\mathcal{H}\left( \chi ;k\right) =\left\{ 
\begin{array}{c}
\sin \chi \ ,\qquad k=1 \\ 
\chi \ ,\qquad k=0 \\ 
\sinh \chi \ ,\qquad k=-1%
\end{array}%
\right. \ ,
\end{equation}%
and 
\begin{equation}
f\left( v,r;k\right) =k-\frac{2m\left( v\right) }{r^{2}}-\frac{\widetilde{%
\kappa }^{2}\widetilde{\Lambda }}{6}r^{2}\ .  \label{ff}
\end{equation}%
Here $\widetilde{\kappa }^{2}=8\pi G_{\left( 5\right) }$ is the
gravitational coupling constant in the bulk ($G_{\left( 5\right) }$
representing the $5$-dimensional gravitational constant) and it has the
dimension of (length$^{4}$mass$^{-1}$time$^{-2}$). The null coordinate $v$
is ingoing (the $v=$ const. lines are outgoing).\ As a simplifying
assumption and because we would like to focus on the radiation effects, we
choose the same bulk cosmological constant $\widetilde{\Lambda }$ on the two
sides of the brane. The mass function $m\left( v\right) $ is freely
specifiable, however different on the left and right sides of the brane: $%
m_{L}\left( v\right) \neq m_{R}\left( v\right) $. \ We note that, as defined
here, the mass function $m\left( v\right) $ has the dimension of length$^{2}$%
. This is, because it denotes $G_{\left( 5\right) }M\left( v\right) /c^{2}$,
with $M\left( v\right) $ the true mass. Since we have chosen $\hbar =1=c$,
dimensionally speaking, length, time and mass$^{-1}$ are the same. Thus
technically the mass function $m\left( v\right) $ represents the cumulated
mass of the bulk black hole and radiation multiplied by $G_{\left( 5\right) }
$\ (of dimension mass$^{-3}$).

The source of the metric (\ref{ChVAdS5}) is pure radiation considered in the
geometrical optics limit (null dust), with energy-momentum tensor 
\begin{equation}
\widetilde{T}_{ab}^{ND}=\psi \left( v,r\right) l_{a}l_{b}\ .  \label{TND}
\end{equation}%
Here $\psi \left( v,r\right) $ is the energy density and $l$ is a
(future-oriented) null 1-form: 
\begin{equation}
l=-\frac{dv}{\dot{v}}=n+u\ .  \label{l}
\end{equation}%
Here $n$ is the dual 1-form of the unit normal to the brane and $u$ the dual
1-form of the timelike (negative) unit vector. We define these vectors in
what follows. 
We give the brane by the embedding relations $v=v\left( \tau \right) $ and $%
r=a\left( \tau \right) $, with $\tau $ the cosmological time. Then we choose
the timelike unit vector $u$ such that $\tau $ is adapted to $u$: 
\begin{equation}
u=\frac{\partial }{\partial \tau }=\dot{v}\frac{\partial }{\partial v}+\dot{r%
}\frac{\partial }{\partial r}\ .
\end{equation}%
(A dot denotes derivatives with respect to $\tau $). As consequence of $%
u^{a}u_{a}=-1$, we have on the brane 
\begin{equation}
f\dot{v}=-\dot{a}+\left( \dot{a}^{2}+f\right) ^{1/2}\ .  \label{vdot}
\end{equation}%
Here we have chosen the positive root in order to have $\dot{v}$ positive.
Now we can write the 1-form field $u_{a}$ as 
\begin{equation}
u=-\left( \dot{r}^{2}+f\right) ^{1/2}dv-\dot{v}dr=-d\tau \ .  \label{tangent}
\end{equation}%
The unit normal 1-form to the brane can be expressed as 
\begin{equation}
n=-\dot{r}dv+\dot{v}dr\ .  \label{normal}
\end{equation}

The bulk Einstein equation for the metric (\ref{ChVAdS5}) and the source
term 
\begin{equation}
\widetilde{\Pi }_{cd}=-\widetilde{\Lambda }\widetilde{g}_{ab}+\widetilde{T}%
_{ab}^{ND}\ ,
\end{equation}%
establishes a connection between the energy density $\psi $ and the time
derivative of the mass function $\dot{m}=\left( dm/dv\right) \dot{v}$ as 
\begin{equation}
\psi =-\frac{3\dot{m}\dot{v}}{\widetilde{\kappa }^{2}a^{3}}\ ,  \label{psi}
\end{equation}%
where $\dot{v}$ is given by Eq. (\ref{vdot}).

The square root from $\dot{v}$ can be eliminated by using the relations
derived in \cite{Decomp}: 
\begin{eqnarray}
\left( \dot{a}^{2}+f_{R,L}\right) ^{1/2} &=&\pm \overline{B}+\frac{\Delta B}{%
2}\ ,  \notag \\
\overline{B} &=&-\frac{\widetilde{\kappa }^{2}a}{6}\left( \rho +\lambda
\right) \ ,  \notag \\
\Delta B &=&\frac{6\Delta m}{\widetilde{\kappa }^{2}a^{3}\left( \rho
+\lambda \right) }\ .  \label{root}
\end{eqnarray}%
(In the above and the forthcoming formulae the first subscript refers to the
upper sign and $\Delta $ always is understood as the difference taken
between the $R$ and $L$ regions. An overbar denotes the average of a
physical quantity, taken on the two sides of the brane.) Therefore Eqs. (\ref%
{vdot}), (\ref{psi}) and (\ref{root}) imply 
\begin{equation}
\psi _{R,L}=-\frac{3\dot{m}_{R,L}}{\widetilde{\kappa }^{2}a^{3}f_{R,L}}\left[
-\dot{a}\mp \frac{\widetilde{\kappa }^{2}a}{6}\left( \rho +\lambda \right) +%
\frac{3\Delta m}{\widetilde{\kappa }^{2}a^{3}\left( \rho +\lambda \right) }%
\right] \ .  \label{psidetailed}
\end{equation}%
Why do we need such a detailed expression for $\psi _{R,L}$? The answer is
that radiation energy density appears in two of the key equations governing
cosmological evolution. (These equations emerge from the modified Einstein
equation.) The energy densities $\psi _{R,L}$ are present both in the
energy-balance equation 
\begin{equation}
\dot{\rho}+3\frac{\dot{a}}{a}\left( \rho +p\right) =-\Delta \psi \ ,
\label{noconserv}
\end{equation}%
and in the Raychaudhuri equation: 
\begin{gather}
\frac{\ddot{a}}{a}=\frac{\Lambda _{0}}{3}-\frac{\kappa ^{2}}{6}\left[ \rho
\left( 1+\frac{2\rho }{\lambda }\right) +3p\left( 1+\frac{\rho }{\lambda }%
\right) \right]   \notag \\
-\frac{2\overline{m}}{a^{4}}+\frac{27\left( p-\lambda \right) \left( \Delta
m\right) ^{2}}{\widetilde{\kappa }^{4}a^{8}\left( \rho +\lambda \right) ^{3}}
\notag \\
-\frac{\widetilde{\kappa }^{2}\overline{\psi }}{3}-\frac{3\Delta m\Delta
\psi }{\widetilde{\kappa }^{2}a^{4}\left( \rho +\lambda \right) ^{2}}\ .
\label{Raychaudhuri}
\end{gather}%
However, the third essential equation, the Friedmann equation is independent
of $\psi $: 
\begin{eqnarray}
\frac{\dot{a}^{2}+k}{a^{2}} &=&\frac{\Lambda _{0}}{3}+\frac{\kappa ^{2}\rho 
}{3}\left( 1+\frac{\rho }{2\lambda }\right)   \notag \\
&&+\frac{2\overline{m}}{a^{4}}+\frac{9\left( \Delta m\right) ^{2}}{%
\widetilde{\kappa }^{4}a^{8}\left( \rho +\lambda \right) ^{2}}\ ,
\label{Friedmann}
\end{eqnarray}%
(The above equations are particular cases of the equations derived in \cite%
{Decomp} in a more generic set-up.) In the above two equations $\Lambda _{0}$
denotes the 4-dimensional cosmological constant given in terms of the brane
tension $\lambda $ and bulk cosmological constant as: 
\begin{equation}
2\Lambda _{0}=\kappa ^{2}\lambda +\widetilde{\kappa }^{2}\widetilde{\Lambda }%
\ .
\end{equation}

The system of equations (\ref{psidetailed})-(\ref{Friedmann}) governs the
motion of the brane, seen from the brane point of view as cosmological
evolution.

\section{The free functions of the model}

There are a number of free functions in our model. Among them, by
straightforward algebra on Eq. (\ref{psidetailed}) the quantities $\Delta
\psi $ and $\overline{\psi }$ can be expressed in terms of $\overline{m}$, $%
\Delta m$, $\overset{\cdot }{\overline{m}}=\overline{\dot{m}}$ and $%
\skew{40}{\dot}{(\Delta m)}=\Delta \dot{m}$ as 
\begin{eqnarray}
\Delta \psi &=&\frac{3}{2\tilde{\kappa}^{2}a^{3}}\left[ \dot{a}-\frac{%
3\Delta m}{\widetilde{\kappa }^{2}a^{3}\left( \rho +\lambda \right) }\right]
\left( 2\overline{\dot{m}}F_{-}+\Delta \dot{m}F_{+}\right)  \notag \\
&&+\frac{1}{4a^{2}}\left( \rho +\lambda \right) \left( 2\overline{\dot{m}}%
F_{+}+\Delta \dot{m}F_{-}\right) \ , \\
\overline{\psi } &=&\frac{3}{4\widetilde{\kappa }^{2}a^{3}}\text{ }\left[ 
\dot{a}-\frac{3\Delta m}{\widetilde{\kappa }^{2}a^{3}\left( \rho +\lambda
\right) }\right] \left( 2\overline{\dot{m}}F_{+}+\Delta \dot{m}F_{-}\right) 
\notag \\
&&+\frac{1}{8a^{2}}\left( \rho +\lambda \right) \left( 2\overline{\dot{m}}%
F_{-}+\Delta \dot{m}F_{+}\right) \text{ },
\end{eqnarray}%
where we have introduced the shorthand notations 
\begin{eqnarray}
F_{-} &=&\frac{1}{f_{R}}-\frac{1}{f_{L}}=\frac{2a^{2}\Delta m}{\left( a^{2}%
\overline{f}\right) ^{2}-\left( \Delta m\right) ^{2}}\text{\ }, \\
F_{+} &=&\frac{1}{f_{R}}+\frac{1}{f_{L}}=\frac{2a^{4}\overline{f}}{\left(
a^{2}\overline{f}\right) ^{2}-\left( \Delta m\right) ^{2}}\text{\ }, \\
a^{2}\overline{f} &=&\left( \frac{\kappa ^{2}\lambda }{6}-\frac{\Lambda _{0}%
}{3}\right) a^{4}+ka^{2}-2\overline{m}\text{\ }.
\end{eqnarray}%
Then in Eqs. (\ref{noconserv})-(\ref{Friedmann}) the role of the free
functions $\overline{\psi }$ and $\Delta \psi $ is taken by $\overline{\dot{m%
}}$ and $\Delta \dot{m}$ .

In what follows, we discuss the physical interpretation of these free
functions. First we show that $\Delta m$ is determined by three quantities.
These are the energy density of the brane, the scale factor and Hubble
parameter. Second, that $\psi _{L}$ is given by the energy emission of the
bulk black-hole, evaluated at the left side of the brane. Finally we
interpret $\Delta \dot{m}$ as the energy absorption on the brane.

\subsection{The energy content of the brane}

By rearranging the Friedmann equation, we find $\Delta m$ as function of the
cosmological perfect fluid energy density $\rho $, scale factor $a$ and
Hubble parameter $H=\dot{a}/a$: 
\begin{equation}
\left( \Delta m\right) ^{2}=\frac{2\kappa ^{2}}{3\lambda }a^{8}\left( \rho
+\lambda \right) \left[ H^{2}+\frac{\overline{f}}{a^{2}}-\frac{\kappa ^{2}}{%
6\lambda }\left( \rho +\lambda \right) ^{2}\right] \ .
\end{equation}%
In the derivation we have employed the relation among the brane tension and
the gravitational coupling constants in $5$ and $4$-dimensions:%
\begin{equation}
\frac{\kappa ^{2}}{\lambda }=\frac{\widetilde{\kappa }^{4}}{6}\ .
\end{equation}%
It is easy to argue, that $m_{R}>m_{L}$ as $m_{R}$ contains in addition to $%
m_{L}$ the contribution from the brane energy momentum. Therefore $\Delta
m>0 $.

We stress here that $\Delta m$ is not strictly the mass of the brane.
Rather, it represents the difference of the mass functions of the VAdS5
metric, evaluated on the two sides of the brane. Although, dimensionally
speaking, $\Delta m$ is mass$^{-2}$, it is directly proportional to the
difference in the cumulated masses $M\left( v\right) $ of the bulk black
hole and radiation on the two sides of the brane.\ The reason why $m_{R}\neq
m_{L}$ is that the brane contributes toward the mass function both though
its energy momentum and its curvature. In a less strict sense then $\Delta m$
can be regarded as a measure of the energy content of the brane.

\subsection{Hawking radiation in the bulk}

The rate of decrease in mass of an evaporating black hole in an SAdS5
spacetime with $k=1$ was computed in \cite{EHM}-\cite{GCL}. In \cite%
{Jennings} the assumption was advanced that this agrees with the rate of
decrease of the mass function $m_{L}\left( v\right) $ in the VAdS5
spacetime, even for arbitrary $k$. Numerical results were given in \cite%
{Jennings} for $k=0$. We adopt here the same identification, but only for $%
k=1$, for which the proof of \cite{EHM}-\cite{GCL} holds. Then the energy
density of the radiation escaping from the black hole, evaluated at the left
side of the brane is: 
\begin{equation}
\psi _{L}=\frac{5\zeta \left( 5\right) }{24\pi ^{9}a^{3}m_{L}\left[ \dot{a}%
+\left( \dot{a}^{2}+f_{L}\right) ^{1/2}\right] }\text{ \ },  \label{Hawking}
\end{equation}%
where $\zeta $ is the Riemann-zeta function. Here $m_{L}$ contains both the
mass of the black hole and the energy of the Hawking radiation. This
radiation spreads away with the velocity of light and it overtakes the
brane, which has a subluminal motion, therefore $\dot{m}_{R,L}<0$. Moreover,
as part of the energy radiated away is captured on the brane, there is less
radiation escaping in the $R$ region than there would be in the absence of
the brane. Therefore $\dot{m}_{L}<\dot{m}_{R}<0$ and in consequence $\Delta 
\dot{m}>0$. For the same reason $\Delta \psi <0$ holds.

\subsection{Energy absorption on the brane}

The decomposition of the pure radiation energy-momentum tensor (\ref{TND})
with respect to the brane can be done by employing Eq. (\ref{l}). We obtain: 
\begin{equation}
\left( \widetilde{T}_{R,L}^{ND}\right) _{ab}=\psi _{R,L}\left(
u_{a}u_{b}+2u_{(a}n_{b)}+n_{a}n_{b}\right) \ .
\end{equation}%
On the $L$ region the radiation escaping the bulk black hole is
characterized by $\psi _{L}$. Some of this radiation will pass unaffected
through the brane into the $R$ region ($\psi _{R}$). Then we have to
interpret the difference of the transmitted and incident energy-momentum
tensors: 
\begin{gather}
\Delta \widetilde{T}_{ab}^{ND}=\Delta \psi \left(
u_{a}u_{b}+2u_{(a}n_{b)}+n_{a}n_{b}\right)  \notag \\
=-\left( \rho ^{rad}u_{a}u_{b}+2q_{(a}^{rad}u_{b)}+p^{rad}n_{a}n_{b}\right)
\delta \left( y\right) \ .  \label{dark}
\end{gather}%
Therefore the difference in the energy density of the radiation arriving
from left and escaping into the right regions appears on the brane as a dust
with energy density $\rho ^{rad}\sim \left( -\Delta \psi \right) >0$. The
energy flux ariving from the bulk into the brane is $q_{a}^{rad}\sim \left(
-\Delta \psi \right) n_{a}$. Finally, the radiation pressure $p^{rad}\sim
\left( -\Delta \psi \right) >0$ drives the brane into an accelerated motion.
This in turn accelerates the cosmic expansion.

With $\Delta \psi <0$, the acceleration caused by the radiation pressure is
manifest from the last term of the Raychaudhuri equation (\ref{Raychaudhuri}%
).

\section{Total absorption on the brane}

In this section we consider a simple model, in which $k=1$ and $\Lambda
_{0}=0$. We allow the brane to absorb \textit{all} of the radiation escaping
from the bulk black hole. With this choice the equations simplify
considerably and we can study the effects of the absorption, which were not
taken into account in previous treatments.

\subsection{Brane dynamics}

For total absorption $\psi _{R}=0$ and $m_{R}$ becomes a constant. Two free
functions are then left, which can be chosen as $m:=\Delta m$ and $\psi
:=\psi _{L}$. The energy-balance and Raychaudhuri equations (\ref{noconserv}%
) and (\ref{Raychaudhuri}) become 
\begin{eqnarray}
0 &=&\dot{\rho}+3\frac{\dot{a}}{a}\left( \rho +p\right) -\psi \ ,
\label{energybalance} \\
\frac{\ddot{a}}{a} &=&-\frac{\kappa ^{2}}{6}\left[ \rho \left( 1+\frac{2\rho 
}{\lambda }\right) +3p\left( 1+\frac{\rho }{\lambda }\right) \right]  \notag
\\
&&-\frac{2m_{R}}{a^{4}}+\frac{m}{a^{4}}+\frac{27\left( p-\lambda \right)
\left( m\right) ^{2}}{\widetilde{\kappa }^{4}a^{8}\left( \rho +\lambda
\right) ^{3}}  \notag \\
&&-\frac{\widetilde{\kappa }^{2}\psi }{6}+\frac{3\psi m}{\widetilde{\kappa }%
^{2}a^{4}\left( \rho +\lambda \right) ^{2}}\ .  \label{Ray}
\end{eqnarray}%
We remark that with the assumption $\psi _{R}=0$ both $\overline{\psi }$ and 
$\Delta \psi $ become determined. The Friedmann equation (\ref{Friedmann})
emerges now as a consequence of the energy-balance equation (\ref%
{energybalance}) and the Raychaudhuri equation (\ref{Ray}). Stated
otherwise, only two of the Friedmann, Raychaudhuri and energy-balance
equations are now independent. (The interdependence of these three equations
is a generic feature of the standard cosmological models, but not of
brane-world universes. Indeed, prior to assuming the total absorption, the
three equations were independent.)

From Eqs. (\ref{vdot}), (\ref{psi}) and (\ref{Hawking}) the variation in
time of the brane mass function is found: 
\begin{equation}
\dot{m}=\frac{5\zeta \left( 5\right) \widetilde{\kappa }^{2}}{72\pi
^{9}\left( m_{R}-m\right) }\text{ }.
\end{equation}%
This equation is an ordinary differential equation which can be integrated
for $t>t_{0}$. The result is:%
\begin{equation}
m=m_{R}-\sqrt{\left( m_{R}-m_{0}\right) ^{2}-\frac{5\zeta \left( 5\right) 
\widetilde{\kappa }^{2}}{36\pi ^{9}}\left( t-t_{0}\right) }\text{ },
\label{mint}
\end{equation}%
where by $m_{0}$ we have denoted the brane mass function before the Hawking
radiation reaches the brane (this happens at $t=t_{0}$)$.$ The brane mass
function $m$ is monotonically increasing in time until $t_{C}=t_{0}+36\pi
^{9}\left( m_{R}-m_{0}\right) ^{2}/5\zeta \left( 5\right) \widetilde{\kappa }%
^{2}$, when it reaches its maximal value $m=m_{R}.$ When this happens, the
bulk black hole has already completely evaporated and the resulting Hawking
radiation is entirely absorbed by the brane.

The square root in Eq. (\ref{Hawking}) can be eliminated by employing Eq. (%
\ref{root}). We obtain 
\begin{equation}
\psi =\frac{5\zeta \left( 5\right) }{24\pi ^{9}a^{3}\left( m_{R}-m\right) %
\left[ \dot{a}+\frac{\widetilde{\kappa }^{2}a}{6}\left( \rho +\lambda
\right) +\frac{3m}{\widetilde{\kappa }^{2}a^{3}\left( \rho +\lambda \right) }%
\right] }\text{ \ }.  \label{Psi2}
\end{equation}%
With this both free functions $m$ and $\psi $ are given explicitly as
functions of time. This has been achieved by specifying both the rate of
absorption and the energy density of the black hole radiation Eq. (\ref%
{Hawking}).

\subsection{Dimensionless variables}

We introduce dimensionless variables in a slightly different way than in %
\cite{LSR} and \cite{RadBrane}. The reason for this is that the chosen
comoving radial coordinate $\chi $ has to be dimensionless in the $k=\pm 1$
cases. Therefore the scale factor has to carry the dimension, unless in the
case $k=0$, discussed in \cite{LSR} and \cite{RadBrane}. The dimensionless
variables introduced below are well suited for $k=\pm 1$ but they also apply
for $k=0$ if in this latter case the scale factor, rather then the $\chi $%
-coordinate is regarded dimensional. These variables are: 
\begin{eqnarray}
\widehat{t} &=&Ct\text{ },\text{\ }\widehat{a}=Ca\text{ },\text{\ }\widehat{H%
}=\frac{H}{C}\text{ },  \notag \\
\text{ }\widehat{\rho } &=&\frac{\rho }{\lambda }\text{ },\text{ }\widehat{p}%
=\frac{p}{\lambda }\text{ },\text{ }\widehat{\psi }=\frac{\psi }{C\lambda }%
\text{ },  \notag \\
\text{ \ }\widehat{m} &=&C^{2}m\text{ \ },\text{ }\widehat{m}_{R}=C^{2}m_{R}%
\text{ }.  \label{dimensionless}
\end{eqnarray}%
Here $C=\kappa \sqrt{\lambda }$ and its inverse represents a distance scale.
(The dimensions of $\kappa ^{2}$ and $\lambda $ are length/mass and
mass/length$^{3}$, respectively.) We note here that the dimension of $\psi $
given by Eqs. (\ref{psi}) or (\ref{Hawking}) is consistent only by choosing $%
\hbar =1=c$, implying that the dimension of mass is length$^{-1}$. Then $C$
can be thought as a mass scale as well. Alternatively, by the choice $c=1$
(therefore length and time having the same dimension), $C$ becomes a basic
energy scale$.$

Written in the variables (\ref{dimensionless}), the system of equations
describing cosmological evolution contains the following dimensionless
parameter: $\widetilde{\kappa }^{8}\lambda ^{3}=36\kappa ^{4}\lambda
=2304\pi ^{2}\lambda /M_{p}^{4}$, where $M_{p}$ denotes the 4-dimensional
Planck mass. By denoting the brane tension as $\lambda =M_{T}^{4}$, the free
parameter can be chosen as $M_{T}/M_{p}$. However for later notational
convenience we introduce the dimensionless parameter $\beta =40\zeta \left(
5\right) \left( M_{T}/M_{P}\right) ^{4}/\left( 3\pi ^{7}\right) $. The
energy-balance and Raychaudhuri equations (\ref{noconserv}) and (\ref%
{Raychaudhuri}), written in terms of the dimensionless variables (\ref%
{dimensionless}) and parameter $\beta $ read%
\begin{eqnarray}
0 &=&\widehat{\rho }^{^{\prime }}+3\widehat{H}\left( \widehat{\rho }+%
\widehat{p}\right) -\widehat{\psi }\ , \\
\widehat{H}^{^{\prime }} &=&-\widehat{H}^{2}-\frac{1}{6}\widehat{\rho }%
\left( 1+2\widehat{\rho }\right) -\frac{1}{2}\widehat{p}\left( 1+\widehat{%
\rho }\right)   \notag \\
&&-\frac{2\widehat{m}_{R}}{\widehat{a}^{4}}+\frac{\widehat{m}}{\widehat{a}%
^{4}}+\frac{9\left( \widehat{p}-1\right) \left( \widehat{m}\right) ^{2}}{2%
\widehat{a}^{8}\left( \widehat{\rho }+1\right) ^{3}}  \notag \\
&&-\frac{\widehat{\psi }}{\sqrt{6}}+\frac{3\widehat{m}\widehat{\psi }}{\sqrt{%
6}\widehat{a}^{4}\left( \widehat{\rho }+1\right) ^{2}}\ ,
\end{eqnarray}%
with 
\begin{equation}
\widehat{\psi }=\frac{\beta }{\widehat{a}^{4}\left( \widehat{m}_{R}-\widehat{%
m}\right) \left[ \widehat{H}+\frac{\widehat{\rho }+1}{\sqrt{6}}+\frac{3%
\widehat{m}}{\sqrt{6}\widehat{a}^{4}\left( \widehat{\rho }+1\right) }\right] 
}  \label{Psihat}
\end{equation}%
and 
\begin{equation}
\widehat{m}^{^{\prime }}=\frac{2\beta }{\sqrt{6}\left( \widehat{m}_{R}-%
\widehat{m}\right) }\text{ }.  \label{mhatdot}
\end{equation}%
Here a prime denotes derivates with respect to the dimensionless time $%
\widehat{t}$ . Eq. (\ref{mhatdot}) integrated gives the temporal evolution
of the dimensionless mass function of the brane 
\begin{equation}
\widehat{m}=\widehat{m}_{R}-\sqrt{\left( \widehat{m}_{R}-\widehat{m}%
_{0}\right) ^{2}-\frac{4\beta }{\sqrt{6}}\left( \widehat{t}-\widehat{t}%
_{0}\right) }\text{ }.  \label{mbhat}
\end{equation}%
The generalized Friedmann equation (\ref{Friedmann}) in dimensionless
variables readily follows: 
\begin{eqnarray}
\widehat{H}^{2} &=&-\frac{1}{\widehat{a}^{2}}+\frac{\widehat{\rho }}{3}%
\left( 1+\frac{\widehat{\rho }}{2}\right) +\frac{2\widehat{m}_{R}}{\widehat{a%
}^{4}}  \notag \\
&&-\frac{\widehat{m}}{\widehat{a}^{4}}+\frac{3\left( \widehat{m}\right) ^{2}%
}{2\widehat{a}^{8}\left( \widehat{\rho }+1\right) ^{2}}\ .
\label{Fried_adimensional}
\end{eqnarray}

\section{Numerical results}

In this section we give numerical results on the evolution of the
brane-world universe in the presence of one radiating bulk black hole. We
assume the movement of the brane starts on the apparent horizon%
\begin{equation}
\widehat{r}_{AH}=\sqrt{-3+\sqrt{9+12\widehat{m}_{L}}}\text{ }
\end{equation}%
of the bulk black hole (therefore $t_{0}=0$). This horizon is dynamical, as $%
\widehat{m}_{L}$ decreases due to the escaping Hawking radiation. The brane
leaves the horizon at $\widehat{m}_{L0}=\widehat{m}_{R}-\widehat{m}_{0}$. In
order to have clear graphical solutions, we have found convenient to chose
the dimensionless mass values of $\widehat{m}_{R}=0.3$ and $\widehat{m}%
_{0}=0.29$. Then the initial position of the apparent horizon is at $%
0.1411870178$. Away from the horizon, the brane absorbes the Hawking
radiation. We investigate a radiation dominated universe, so $\widehat{p}=%
\widehat{\rho }/3$. We choose the free parameter as $M_{T}/M_{P}=1/10$. In
what follows, we compare the evolutions of the brane in the absence and in
the presence of the radiation coming from the bulk black hole.

\subsection{The evolution of the brane-world without bulk black hole
radiation}

In the simplest case when we switch off the radiation from the bulk black
hole, beyond the standard density and curvature source terms in the
Friedmann equation (\ref{Fried_adimensional}), the model includes dark
radiation and a strong asymmetry due to the chosen setup of a single left
black hole in the bulk. We would like to see the evolution of such a brane
universe, for later comparison with the radiating case. With the Hawking
radiation from the bulk black hole switched off, $\widehat{m}=$const.

\begin{figure}[tbp]
\begin{center}
\includegraphics[height=6cm]{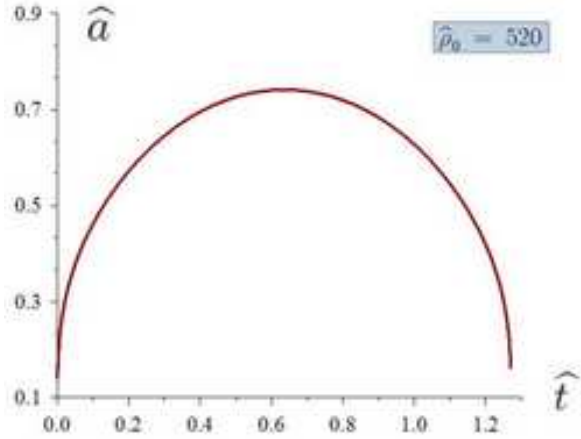}
\end{center}
\caption{The evolution of the scalefactor in the absence of the Hawking
radiation shows the characteristic pattern of a closed universe. The plot is
for the initial brane density $\widehat{\protect\rho }_{0}=520$.}
\label{Fig2}
\end{figure}

\begin{figure}[tbp]
\begin{center}
\includegraphics[height=6cm]{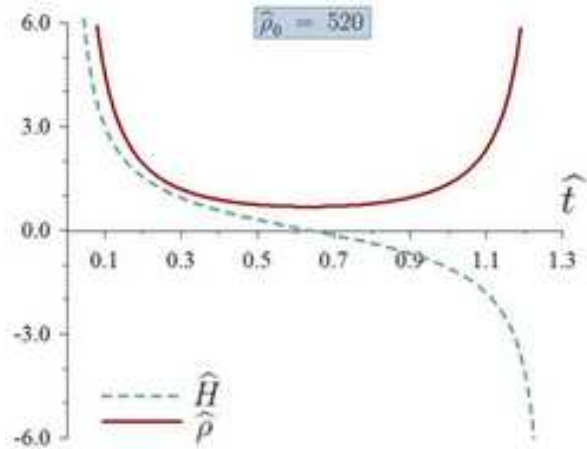}
\end{center}
\caption{Evolution in the absence of the Hawking radiation of the Hubble
parameter (dotted line) and brane energy density (solid line). The initial
brane density is $\widehat{\protect\rho }_{0}=520$.}
\label{Fig3}
\end{figure}

Our numerical analysis shows that none of the brane effects (quadratic
energy density, dark radiation and asymmetry source terms) will change
significantly the usual behaviour of a closed ($k=1$) universe, as can be
seen from the behaviour of the scale-factor, represented in Fig \ref{Fig2}.
The Hubble parameter and the brane energy density evolve accordingly. The
Hubble parameter continuously decreases, while the energy density tends to
infinity both at the beginning and at the end of this brane-world universe.
These behaviours are shown in Fig \ref{Fig3}.\qquad 

\begin{figure}[tbp]
\begin{center}
\includegraphics[height=6cm]{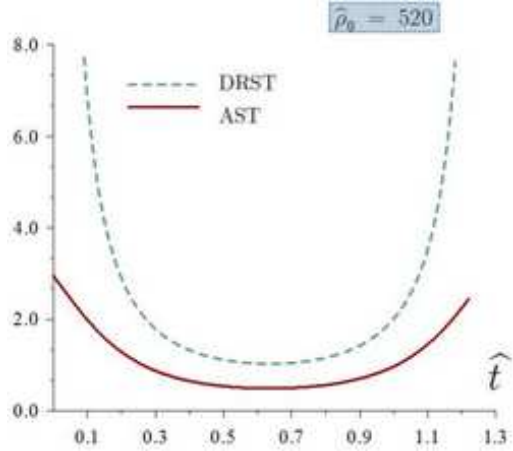}
\end{center}
\caption{Evolution of the dark radiation (DRST, dotted line) and asymmetry
(AST, solid line) source terms of the Friedmann equation (for the initial
brane density $\widehat{\protect\rho }_{0}=520$).}
\label{Fig4}
\end{figure}

\begin{figure}[tbp]
\begin{center}
\includegraphics[height=6cm]{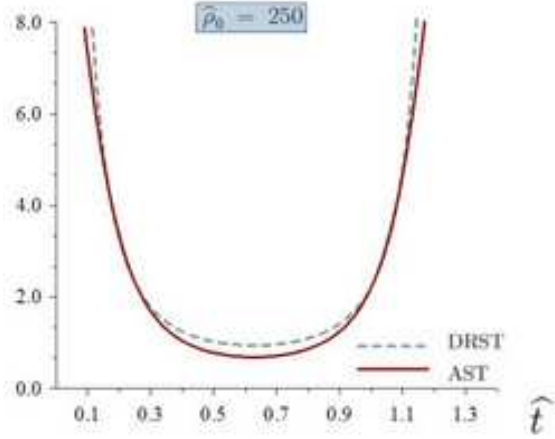}
\end{center}
\caption{Similar evolutions of the dark radiation (dotted line) and
asymmetry (solid line) source terms in the Friedmann equation at $\widehat{%
\protect\rho }_{0}=250$.}
\label{Fig5}
\end{figure}

\begin{figure}[tbp]
\begin{center}
\includegraphics[height=6cm]{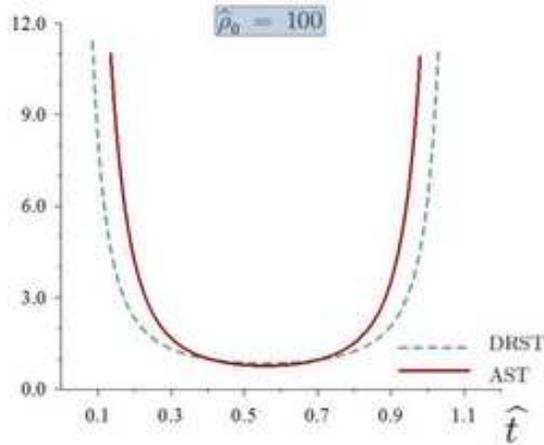}
\end{center}
\caption{For light branes ($\widehat{\protect\rho }_{0}=100$) the asymmetry
(solid line) source term dominates over the dark radiation (dotted line)
source term in the Friedmann equation for both early and late universes.}
\label{Fig6}
\end{figure}

The evolutions of the dark radiation source term%
\begin{equation}
\frac{2\widehat{m}_{R}}{\widehat{a}^{4}}-\frac{\widehat{m}}{\widehat{a}^{4}}%
\ ,
\end{equation}%
and of the asymmetry source term%
\begin{equation}
\frac{3\left( \widehat{m}\right) ^{2}}{2\widehat{a}^{8}\left( \widehat{\rho }%
+1\right) ^{2}}\ 
\end{equation}%
are represented in Fig \ref{Fig4}. Both contributions are positive, so they
act in a similar way to ordinary matter sources in the Friedmann equation.
The dark radiation dominates over the asymmetry during the whole evolution,
the difference being less when the universe reaches its maximal size. 
As the initial value of the brane energy density is lowered, these two
source terms become comparable (Fig \ref{Fig5}). For even lower initial
energy densities the asymmetry source term dominates at both early and late
times (Fig \ref{Fig6}). 

\subsection{The energy density of the Hawking radiation}

\begin{figure}[tbp]
\begin{center}
\includegraphics[height=6cm]{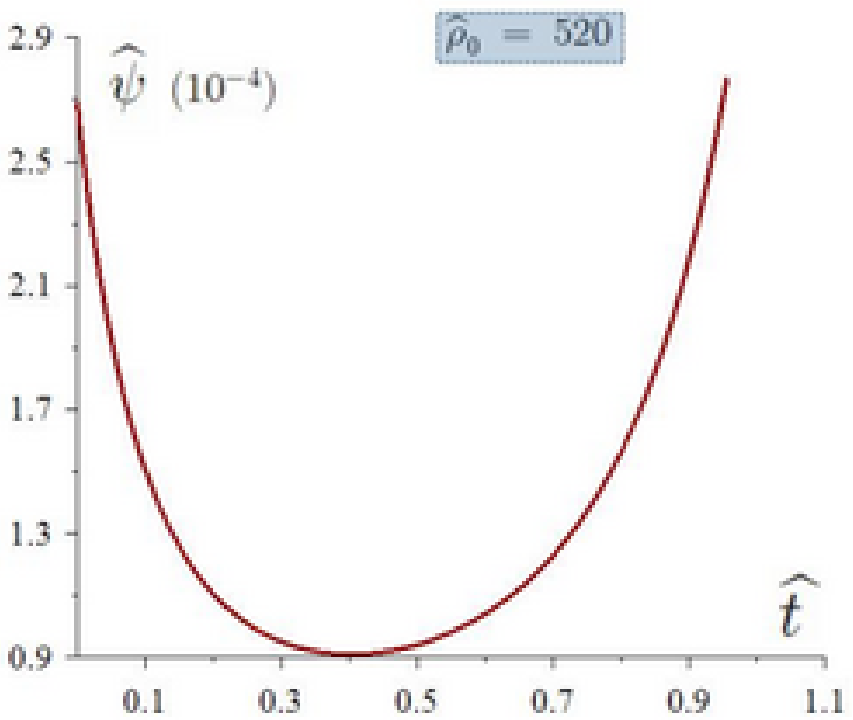}
\end{center}
\caption{The evolution of the energy density of Hawking radiation for $%
\widehat{\protect\rho }_{0}=520$.}
\label{Fig7}
\end{figure}

The energy density $\widehat{\psi }$ (evaluated near the brane, on the left
region) of the Hawking radiation from the bulk black hole evolves; cf Fig %
\ref{Fig7}. 
In the expanding era $\widehat{\psi }$ shows a sharp decrease. This is due
to the fact that the distance between the brane and the horizon increases.
Interestingly $\widehat{\psi }$ starts to increase before the maximal size
of the closed universe is reached. Later on $\widehat{\psi }$ keeps
increasing as the universe contracts. The higher values of $\widehat{\psi }$
towards the end of contraction mathematically can be explained by the
presence of the forever decreasing factor $\widehat{H}$ in the denominator
of $\widehat{\psi }$ (see Eq. (\ref{Psihat})). Stated otherwise, from Eq. (%
\ref{psi}) $\psi \symbol{126}\dot{v}$ so that from a physical point of view,
the increase in $\widehat{\psi }$ at late times, exceeding its value at
early times can be explained as a Doppler effect. The contracting brane
absorbs more energy in unit time than the expanding brane at the same scale
factor value.

We have checked, that with the Hawking radiation switched on, the basic
behaviour of the closed universe will be very similar to that in the case
without this radiation, discussed in the previous subsection. Even in the
presence of the radiation the cosmological evolution shows the same patters.
Although there are modifications, which will be discussed in detail in the
following subsection, they behave as small perturbations during the whole
evolution of this asymmetric brane-world universe.

\subsection{The evolution of the brane-world with the bulk black hole
radiation}

The evolution of the Hubble parameter, the scalefactor, the brane energy
density and different source terms of the Friedmann equation, in the \textit{%
\ presence of the radiation from the bulk black hole} is very similar to
their evolutions in the absence of the radiation. This is due to a delicate
balance between two effects. First, the brane is pushed away by the
radiation pressure, which tends to accelerate it. Second, as all of the
radiation is absorbed on the brane, the self-gravity of the brane increases,
which tends to decelerate it. Still, there are interesting, through small
effects due to the radiation. We illustrate them by representing the \textit{%
differences} in the scalefactor, the brane energy density and different
source terms of the Friedmann equation in the radiating and non-radiating
cases. More specifically, we denote by $A_{\psi }$ the quantity $A$ in the
radiating case and we plot the evolution of $\delta A=A_{\psi }-A$ .

\begin{figure}[tbp]
\begin{center}
\includegraphics[height=6cm]{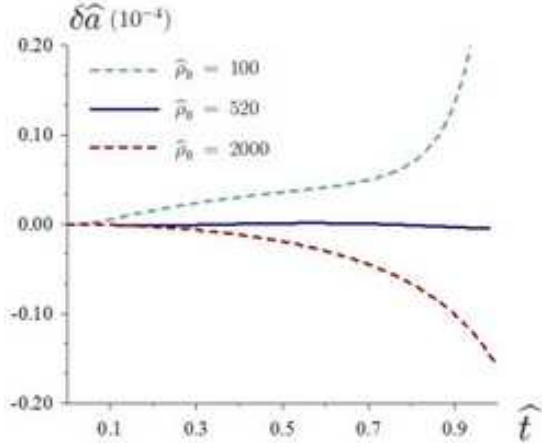}
\end{center}
\caption{The difference between the scale factors in the radiating and
non-radiating case versus time. For $\widehat{\protect\rho }_{0}=520$ a
critical-like behavior is observed (solid line), when the increase of the
self-gravity of the brane due to absorbtion is roughtly compensated by the
radiation pressure. For initial densities much smaller (e.g. $\widehat{%
\protect\rho }_{0}=100$, upper dotted line), the effect of the radiation
pressure is dominant. For higher initial densities than the critical one ($%
\widehat{\protect\rho }_{0}=2000$, lower dotted line) the increase of
self-gravity due to absorbtion dominates over the radiation pressure.}
\label{Fig8}
\end{figure}

In Fig \ref{Fig8} we show the evolution of the differences in the
scalefactors when the radiation is switched on and off, for three different
initial brane energy densities. Surprisingly, we have found a critical-like
behaviour for $\widehat{\rho }_{0}=520$. The difference in the scalefactors
is extremely close to zero, but it shows a sinusoidal-like pattern. For
higher initial densities the scalefactor is smaller in the radiating case
throughout the cosmological evolution. This means that the self-gravity of
the absorbed radiation dominates over the radiation pressure. For initial
pressures higher than the critical one, the scalefactor is higher in the
radiating case during the whole cosmological evolution. This means that the
radiation pressure dominates over the self-gravity of the absorbed
radiation. 

\begin{figure}[tbp]
\begin{center}
\includegraphics[height=6cm] {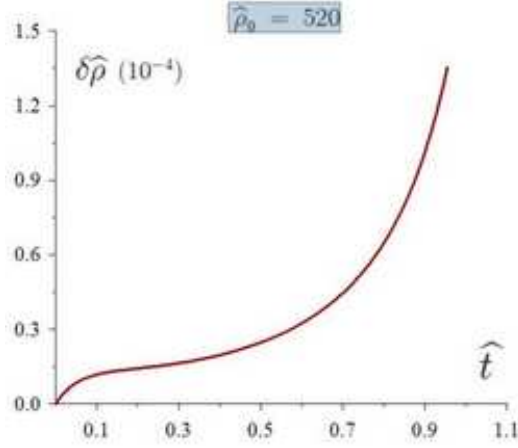}
\end{center}
\caption{The difference between the brane energy densities in the radiating
and non-radiating case for $\widehat{\protect\rho }_{0}=520$.}
\label{Fig9}
\end{figure}

The behaviour of the difference in the energy densities in illustrated in
Fig \ref{Fig9}. Among the previous graphs, Figs \ref{Fig2}, \ref{Fig3}, \ref%
{Fig5} and \ref{Fig9} were plotted for the critical initial density. For
other density values these graphs would be merely scaled, and would show
similar features.

\begin{figure}[tbp]
\begin{center}
\includegraphics[height=6cm]{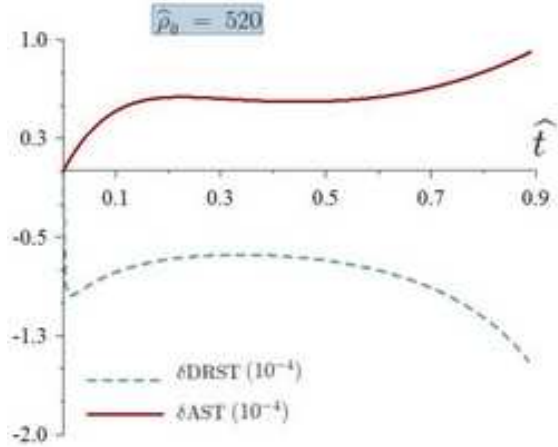}
\end{center}
\caption{The difference between source terms in the Friedmann equation
computed when the radiation is switches on and off. The asymmetry source
term ($\protect\delta $AST, solid line) is increased, while the dark
radiation source term ($\protect\delta $DRST, dotted line) is decreased by
radiation. The plot is for the ''critical'' initial density $\widehat{%
\protect\rho }_{0}=520$. }
\label{Fig10}
\end{figure}

\begin{figure}[tbp]
\begin{center}
\includegraphics[height=6cm]{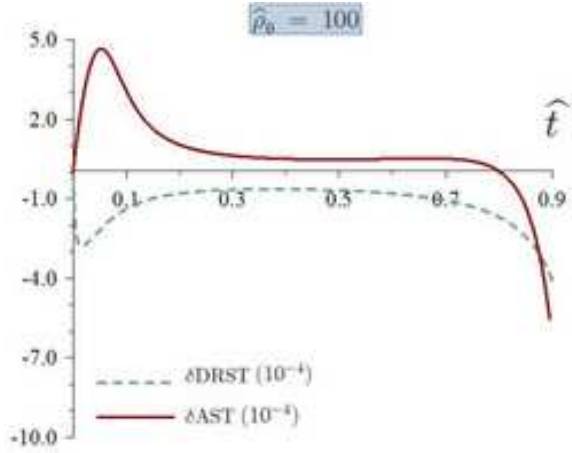}
\end{center}
\caption{The same as in Fig \ref{Fig10}, but for $\widehat{\protect\rho }%
_{0}=100$. The main difference with respect to Fig \ref{Fig10} is that at
early times the increase in the asymmetry term is faster than the decrease
in the dark radiation term.}
\label{Fig11}
\end{figure}

\begin{figure}[tbp]
\begin{center}
\includegraphics[height=6cm]{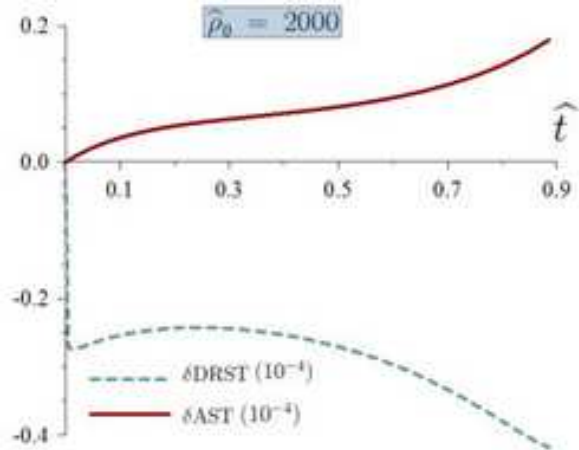}
\end{center}
\caption{The same as in Fig \ref{Fig10}, but for $\widehat{\protect\rho }%
_{0}=2000$. The decrease in the dark radiation term dominates over the
increase of the asymmetry term. This effect is more accentuated than at the
critical initial density, shown on Fig \ref{Fig10}.}
\label{Fig12}
\end{figure}

This is not true for the source terms of the Friedmann equation. We have
seen various behaviours even in the non-radiating case (Figs \ref{Fig4}-\ref%
{Fig6}), when the initial brane energy density is varied. In Figs \ref{Fig10}%
-\ref{Fig12} we have plotted the evolution of the difference, in the
radiating and non-radiating cases, of the asymmetry and dark radiation
source terms, for the critical initial energy density (Fig \ref{Fig10}), for
a much lighter brane (Fig \ref{Fig11}) and for a much heavier brane (Fig \ref%
{Fig12}). With the exceptions of the very early and very late stages of the
evolution of the universe, the two contributions roughly cancel each other
for the critical initial energy density. For light branes the increase in
the asymmetry term at early times is greater than the decrease in the dark
radiation term. For heavy branes, by contrast, the decrease in the dark
radiation term due to the Hawking radiation from the bulk exceeds the
increase of the asymmetry term.

\section{Discussion}

Table-top experiments \cite{Gaccuracy} on possible deviations from Newton's
law currently probe gravity at sub-millimeter scales and as a result they
constrain the characteristic curvature scale of the bulk to $l\lesssim 0.1$
mm. Expressed in units of inverse energy (when $c=1=\hbar $) this curvature
scale is 
\begin{equation}
l_{\max }=506.7\,7\,\text{eV}^{-1}=0.5067\,7\times 10^{12}\,\text{GeV}%
^{-1}\,.
\end{equation}%
This is usually known as $l\lesssim $ $10^{12}\,$GeV$^{-1}$ \cite{MaartensLR}%
.

The $4$-dimensional coupling constant $\kappa ^{2}$ and $4$-dimensional
gravitational constant $G$ are related to the 4-dimensional Planck mass $%
M_{P}$ as $\kappa ^{2}=8\pi G=8\pi /M_{P}^{2}$, with $M_{P}\approx 10^{19}\,$%
GeV$.$ The $5$-dimensional Planck mass is defined as $\widetilde{\kappa }%
^{2}=8\pi /M_{\left( 5\right) }^{3}$. As $M_{\left( 5\right) }$ depends on
both $M_{P}$ and on the characteristic curvature scale $l$ as $M_{\left(
5\right) }^{3}=M_{P}^{2}/l$ \cite{RS2}, we get 
\begin{equation}
M_{\left( 5\right) \min }=6.\,\allowbreak 65\times 10^{8}\,\text{GeV}\,.
\end{equation}%
This result is referred usually as $M_{P}>M_{\left( 5\right) }>10^{8}$ GeV %
\cite{MaartensLR}. Due to the presence of the extra dimension accessible via
gravity, brane-world theories allow for the effective $4$-dimensional Planck
scale on the brane to be be much higher than the true, $5$-dimensional
Planck scale.

The knowledge of both $l$ and $M_{P}$ gives a characteristic lower limit for
the brane tension in the two-brane model \cite{RS1}. In units $c=1=\hbar $
this is%
\begin{equation}
\lambda _{\min }=\frac{3M_{P}^{2}}{4\pi l^{2}}=138.59\,\,\text{TeV}^{4}\,.
\end{equation}%
This result is frequently quoted as $\lambda >1$ TeV$^{4}$ \cite{MaartensLR}%
. On the other hand Big Bang Nucleosynthesis (BBN) constraints result in a
much milder lower limit, $\lambda \gtrsim 1$ MeV$^{4}$ \cite{nucleosynthesis}%
. There is also an astrophysical limit for $\lambda $, this being sensitive
to the equation of state of a neutron star on the brane \cite{GM}. For a
typical neutron star (with uniform density) this is $\lambda >5\,\times
10^{8}$ MeV$^{4}$, in between the two previous lower limits.

Our choice for $M_{T}/M_{P}=1/10$ employed in the numerical analysis of the
previous Section and the numerical value of the Planck-mass $M_{P}\approx
10^{19}\,$GeV implies $M_{T}\approx 10^{15}$TeV. \ In consequence $\lambda
=\left( M_{T}\right) ^{4}=10^{60}$ TeV$^{4}$, well above all previously
enlisted lower bounds. This choice is in agreement with the range of $%
\lambda $ represented in Fig. (1) of Ref. \cite{nucleosynthesis}. The reason
we have chosen such a high value of the brane tension is the following.
Decreasing $\lambda $ would mean to decrease $M_{T}$ accordingly. As $\psi $
is proportional to $\left( M_{T}/M_{P}\right) ^{4}$ (cf Eq. (\ref{Psihat})),
by choosing a significantly smaller value of $\lambda $ the effect of the
radiation energy density becomes very small, in the range of the numerical
errors.

In this paper our intention was to focus on the effects of the asymmetric
setup and bulk radiation absorbed on the brane. Therefore we have simplified
the model by setting $\Lambda =0$. By means of our choices of $\widehat{m}%
_{R}$ and $\widehat{m}_{0}$ the dark radiation term was chosen of comparable
size to the matter energy density source term, slightly above the limits
established in \cite{BDEL} and \cite{BBNconstraint} from BBN for the dark
radiation term. We note however that those limits were derived in a \textit{%
symmetric} setup, and in the presence of a cosmological constant, which is
not the case here.

The question arises of whether the dark radiation term can take the role of
a cosmological constant producing cosmic acceleration. Contrarily to
optimistic expectations, this is not exactly happening in brane-world
theories. Although the dark radiation term $2\overline{m}/a^{4}$ is positive
(for $\overline{m}>0$) in the Friedmann equation (\ref{Friedmann}), it
appears with negative sign in the Raychaudhuri equation (\ref{Raychaudhuri}%
). Therefore dark radiation increases both the cosmic expansion rate and the
cosmic deceleration as compared to standard cosmology. A slower expansion
rate and additional acceleration is gained for negative values of $\overline{%
m}$, but such a possibility is not allowed by our present model.

Would the bulk radiation pushing away the brane and therefore causing cosmic
expansion be able to produce inflation on the brane for any set of numerical
data? For the range of numerical parameters employed in the preceding
Section such an effect does not occur. Without radiation, the scale factor
shows the typical evolution of a closed universe; cf Fig \ref{Fig2}$.$
Inflation would mean the occurrence of an exponential expansion, but this is
not happening by switching on the radiation. Indeed, a glance on the scales
of Figs \ref{Fig2} and \ref{Fig8}, respectively (in the latter we have
represented the differences in the scale factor with and without radiation),
shows that radiation is not changing significantly the evolution of the
scale factor. The deviations of physical quantities evaluated in the
radiating case as compared to the nonradiating case are typically of the
order of $10^{-4}$. We have also checked that the radiation is not changing
significantly the evolution of the Hubble-parameter depicted in Fig \ref%
{Fig3}, therefore inflation due to this mechanism is excluded.

Stated more generically and independently of the specific numerical choices
in our analysis, we can pose the question, whether the absorbed radiation
could behave like a scalar field $\phi $ in a slow-roll regime. Eq. (\ref%
{dark}) shows that the the absorbed radiation is better interpreted as dark
dust matter, rather than dark radiation. Indeed, it is a pressureless fluid.
Therefore, the attempt to describe it in terms of a scalar field would yield 
$\ddot{\phi}\approx 2V$ and the slow-roll condition could not be satisfied.

As we are considering closed cosmological models with $k=1$, it is obvious
that we should ask whether the radiation pressure would be able to stop the
recollapse of the brane-world universe, producing a bounce? Our numerical
investigations suggest a negative answer. For the whole range of initial
brane energy densities that we have checked, the recollapse will inevitably
occur. The explanation is again the double role of the radiation: it gives
rise to both a radiation pressure, which is accelerating the motion of brane
in the bulk, and an increase of brane self-gravity, which is decelerating
it. As consequence, the Hubble parameter is zero only once during the whole
cosmological evolution, when the expansion ceases and the recollapse begins.

Although radiation pressure can in principle accelerate the cosmological
expansion, it will always encounter the competing effect of the brane
becoming heavier as it absorbes more and more radiation. In our model
radiation pressure acts as dark energy during the whole cosmological
evolution; however, dark matter (some form of dust) is produced as well - cf
Eq. (\ref{dark}). This is counterbalancing the accelerating effect.

Our numerical investigations show, that the increase in the energy density $%
\rho ^{rad}$, which is due to the absorbed radiation is not significant as
compared to the energy density of conventional matter. This can be seen from
our Fig \ref{Fig9}, representing the variation in $\rho $ due to the bulk
black hole radiation. The variation is of order $10^{-4}$ as compared to the
energy density of the brane in the non-radiating case, shown in Fig \ref%
{Fig3}. It is important to mention this point, as in our model we have
studied a \textit{radiation dominated} universe. By contrast, all dark
matter born from the Hawking radiation absorbed on the brane is \textit{dust}%
, cf. Eq. (\ref{dark}). As the absorbed energy density merely represents a
small fraction of the original brane energy density, our assumption of a
radiation-dominated universe holds true.

As we have seen, even the tiny variations of order $10^{-4}$ in the physical
quantities can be achieved only at the price of having a huge brane tension $%
\lambda =10^{60}$ TeV$^{4}$. With the Hawking radiation energy density so
small, we do not expect drastical changes to occur in the mass of the bulk
black hole, although theoretically a complete evaporation may occur at $t_{C}
$ (see the remarks following Eq. (\ref{mint})). Indeed, the coefficient of $%
\left( t-t_{0}\right) $ in Eq. (\ref{mint}) is very small, therefore the
mass of the brane will not increase significantly due to the absorbtion of
this radiation, even seen during the whole cosmological evolution.
Accordingly, the bulk black hole will not completely evaporate, rather its
mass function will decrease only slightly during the lifetime of the closed
universe. How small the rate of evaporation of the bulk black hole is can be
better seen from Eq. (\ref{mbhat}), containing dimensionless variables.
There, with our choice $M_{T}/M_{P}=1/10$, the coefficient of $\left( 
\widehat{t}-\widehat{t}_{0}\right) $ can be evaluated to give%
\begin{equation}
\frac{4\beta }{\sqrt{6}}=\frac{160\zeta (5)}{3\sqrt{6}\pi ^{7}}\times
10^{-4}\approx 7.475\times 10^{-7}\text{ }.
\end{equation}%
This means that starting with the initial (dimensionless) mass function of
the bulk black hole of $\widehat{m}_{L}\left( 0\right) =0.01$, towards the
end of the recollapse the mass function decreases to the values presented in
Table \ref{Table1}. We note that he value of the initial brane energy
density has a certain influence on the final value of the dimensionless mass
function, the evaporation being maximal for the critical brane energy
density. 
\begin{table}[h]
\caption{Due to Hawking radiation, the dimensionless mass function of the
bulk black hole slightly decreases during cosmological evolution. Towards
the end of the recollapse (at $\widehat{t}=1.115$) the dimensionless bulk
black hole mass function is minimal for the \textit{critical} value of the
initial brane energy density.}
\label{Table1}%
\begin{equation*}
\begin{tabular}{|c||c|c|c|}
\hline
$\widehat{m}_{L}$ & $\widehat{\rho }_{0}=100$ & $\widehat{\rho }_{0}=520$ & $%
\widehat{\rho }_{0}=2000$ \\ \hline\hline
$\widehat{t}=0$ & $0.01$ & $0.01$ & $0.01$ \\ \hline
$\widehat{t}=1.115\;$ & $0.0099582386$ & $0.0099582385$ & $0.0099582387$ \\ 
\hline
\end{tabular}%
\end{equation*}%
\end{table}

\section{Concluding Remarks}

In this paper we have considered a scenario which is able to deal with the
possibility of a Hawking radiation escaping a bulk black hole and we have
investigated the consequences of such a radiation on the cosmological
evolution of a closed $k=1$ universe.

For this purpose first we have developed a formalism suitable for describing
the asymmetric brane-world containing a single radiating bulk black hole on
the left bulk region. In the most generic scenario the radiation can be
partially absorbed, partially transmitted and partially reflected on the
brane. We have considered here the case of negligible reflection, because we
wanted to dispose of an exact solution of the Einstein equations in the
bulk. By suppressing the reflection, and considering the geometrical optics
limit of the radiation, the bulk can be described by the VAdS5 metric.

The part of radiation absorbed on the brane generates quite interesting
effects, which were not studied before. These can be related to both dark
energy and dark mattter. In order to concentrate on these effects, the
absorption was chosen as maximal. By suppressing transmission, the equations
could be partially integrated, cf. Eq. (\ref{mint}). Then we have introduced
dimensionless quantities in order to perform a complex numerical analysis.

We have carefully compared the cosmological evolutions of the models both
without and with radiation and found that the Hawking radiation of the bulk
black hole represents merely a perturbation of the non-radiating model. This
holds true during the entire cosmological evolution.

We have studied these perturbations in detail and shown that in the presence
of the radiation, there are two competing effects. The absorption on the
brane contributes towards the self-gravity of the brane, which then tends to
recollapse faster. This phenomenon appears as continuously emerging dark
matter on the brane. By contrast, radiation pressure tends to accelerate
away the brane from the bulk black hole, producing an equivalent of dark
energy. For the numerical data employed, we have found the critical value of
the initial brane energy density for which these two competing effects
roughly annihilate each other. We have also analysed cosmological evolution
for both lighter and heavier branes and found that (at early times) the
radiation pressure is dominant for light branes, while for heavy branes the
self-gravity dominates during the whole cosmological evolution.

As a main result, we have proved that the asymmetry introduced in the model
is not able to change the recollapse of the universe in the $k=1$ case,
regardless of whether the radiation is switched off or on. Our numerical
analysis has shown that the bulk Hawking radiation cannot change the final
fate of the recollapsing universe either.

\section{Acknowledgments}

We thank Roy Maartens for useful comments and for stimulating interactions,
and Ibolya K\'{e}p\'{\i}r\'{o} for assistance in producing the graphs. This
work was supported by OTKA grants no. T046939 and TS044665. L\'{A}G wishes
to thank the support of the J\'{a}nos Bolyai Scholarship of the Hungarian
Academy of Sciences.

\end{document}